# Subcoercive and multilevel ferroelastic remnant states with resistive readout

B. Kundys,[+] V. Iurchuk, C. Meny, H. Majjad, and B. Doudin

*Institut de Physique et Chimie des Matériaux de Strasbourg (IPCMS), UMR 7504 CNRS-UdS, 23 rue du Loess, 67034 Strasbourg, France*

Ferroelectric devices use their electric polarization ferroic order as the switching and storage physical quantity for memory applications. However, additional built-in physical quantities and memory paradigms are requested for applications. We propose here to take advantage of the multiferroic properties of ferroelectrics, using ferroelasticity to create a remnant strain, persisting after stressing the material by converse piezoelectricity means. While large electric fields are needed to switch the polarization, here writing occurs at subcoercive much lower field values, which can efficiently imprint multiple remnant strain states. A proof-of-principle cell, with the simplest and non-optimized resistance strain detection design, is shown here to exhibit 13-memory states of high reproducibility and reliability. The related advantages in lower power consumption and limited device fatigue make our approach relevant for applications.

http://dx.doi.org/10.1063/1.4883375

Ferroelectric (FE) memory effect is based on a remnant net polarization phenomenon, with the electric dipoles orientation modified under the application of an electric field. Although this process is always accompanied with a net deformation, the strain has not been deeply explored as an alternative parameter exhibiting hysteresis that can be electrically switched and used for storage applications.[1] Elastic deformations are usually considered as a technology bottleneck, plaguing the stability and reproducibility of FE cells. There is however a growing interest in multiferroic materials, exhibiting several ferroic orders sometimes coupled,[2] with possible applications for multifunctional devices. For spin electronics applications, strain is often a common physical property linking magnetic and electrical ferroic orders. Strain is also becoming an increasingly popular method for electrical control of magnetization,[3–10] magnetoresistance,[11] or magnetic domain wall propagation,[12,13] through converse piezoelectric effect of ferroelectric substrates. The recent claims of low power consumption[14,15] in such systems make strain-controlled devices possible contenders for future logic circuit design. There are however only very few reports taking direct advantage of the ferroelastic properties of the materials for multiferroic studies, with the noticeable recent report of using flexoelectric properties,[16] using stress gradient to control ferroelectric domains.[17] One symmetry reason limiting the direct use of ferroelasticity is the fact that the strain value is identical in the two saturated electric polarization states for electric fields exceeding coercive values. There is therefore a need for clarifying subcoercive field strain hysteresis dynamics possibly appearing in many piezoelectric materials, to avoid confusions when interpreting the piezoelectrically controlled magnetization in magnetic thin films.

We show here how tunable bulk strain hysteresis can be realized in a very simple structure at small excitation electric field values on the most common and widely used FE material lead zirconate titanate (PZT)[18] used as a substrate. The strain curve is measured with an improved capacitance dilatometer similar to that described in Ref. 19 and shows typical butterfly-shape curve behavior,[20] reaching a maximum near the FE coercive force (upper inset of Figure 1). Large strains created when poling a ferroelectric material are usually considered as technically unattractive,[21] leading to fatigue and instability of the device as they involve large deformations. By contrast, the subcoercive curve shows much smaller deformation level with a completely different symmetry and shape, revealing a well-defined unsaturated hysteresis loop (Fig. 1). There is therefore a high interest in studying inverse piezoelectric behaviour at subcoercive stress fields, where the available literature is quite limited.

The strain here can alternatively be measured using a stripe of CoFe thin film that changes its resistance in response to the deformation (Fig. 1 (lower inset)). The 150 nm thick CoFe film was deposited in an e-beam evaporator, with base pressure of $4 \times 10^{-8}$ millibars and at a rate of 0.2 nm/s. The electrodes were connected using silver paste and the resistance of the film was measured employing an Agilent LCR meter at 1 kHz frequency along the y axis. The substrate was $0.49 \times 2 \times 0.2$ mm in xyz sizes and the top CoFe line was 0.068 mm wide and 1.3 long. We took advantage of the stripe piezoresistive response that is sensitive to its geometrical parameters change.[22] For the purpose of presenting a proof-of-principle experiment, with straightforward measurable signal, we did not include any design improvement in the stripe material and geometry and did not use a measurement bridge for sensitivity purposes. As shown in Fig. 2, the resistance of the deposited film follows the strain hysteresis as expected. The electric field is applied to an initial state sample with resistance R0, with remnant memory states of resistance values R1 and R2 persisting at subsequent removal of the applied field. Due to the Poisson effect, the compressive deformation along X leads to the elongation along the two other axes. Taking into account the changes in

[+]Author to whom correspondence should be addressed. Electronic mail: kundysATipcms.unistra.fr

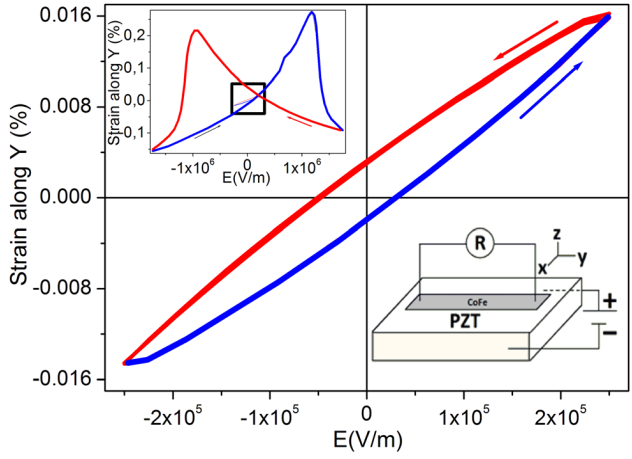

FIG. 1. Subcoercive ferroelastic loop. Upper inset shows high electric field piezoelectric loop with highlighted subcoercive ferroelastic region. Lower inset shows schematic of the experiment with PZT ceramics and film to measure resistance (Fig. 2).

y- and z-dimensions of the film, we can estimate the relative change of the CoFe film resistance. The experimental and computed values are of the same sign and order of magnitude: $dR/R_{calculated} \approx 7.9 \times 10^{-5}$ while $dR/R_{experimental} \approx 6.4 \times 10^{-5}$ for an applied field of about $1.4 \times 10^5$ V/m.

The subcoercive deformation imprinted in the stripe resistance state has a magnitude that depends on the maximum amplitude of the maximum stress voltage, making easy the implementation of a multistate memory element, while keeping the applied electric fields well below the coercive force of the FE polarization. Fig. 2 shows that other remnant resistance values can be obtained when increasing the maximum applied electric field. (R3-R6), creating a multi-state memory cell. The switching is very reproducible upon cycling, with testing of $10^6$ cycles without detecting fatigue. We attribute this stability to the applied electric fields amplitude that is significantly below the FE coercive force, allowing us building many stable memory states by voltage pulse amplitude and sign simple manipulations, avoiding poling the FE. Figure 3 illustrates the stability and diversity of the multiple resistance states created by this technique. The number of presented resistance states values is not a limit, and we recall that the data of Fig. 3 are obtained without optimization of the strain gauge material, geometry, dimensions, and PZT surface quality. Importantly, although it is probably impossible to recover the exact domain configuration of the virgin state, the corresponding deformation equilibrium and therefore the corresponding initial resistance of the film can repeatedly be recovered on-demand by applying a damping voltage with a time profile shown in Figure 3(b). We also observe that a given resistance state can remain stable over weeks, confirming the interest of the low stress fields approach for memory applications.

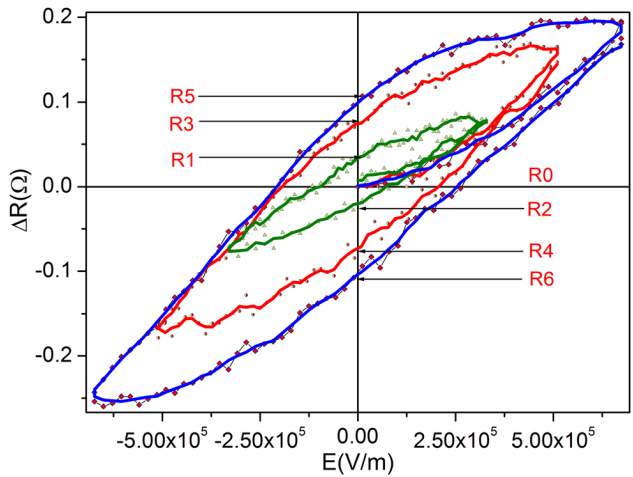

FIG. 2. Electroresistive hysteresis of CoFe film with multi memory states. Solid lines represent smoothed experimental results.

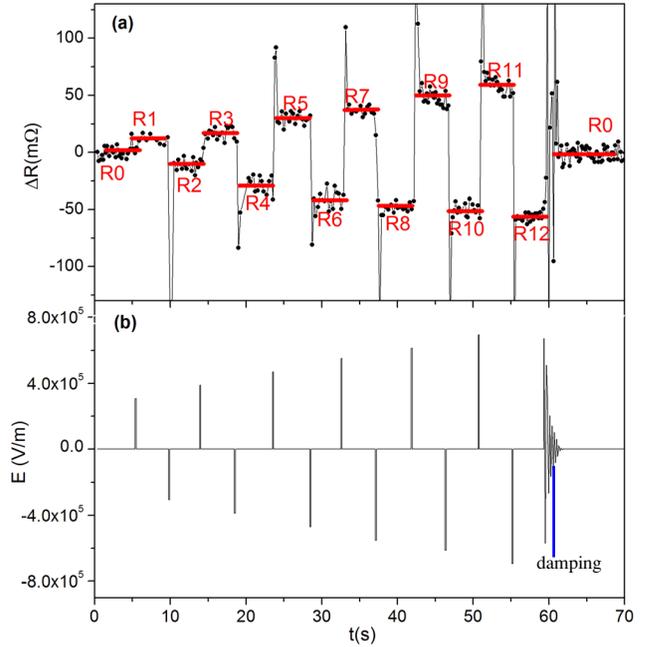

FIG. 3. Electric field induced switching between multiple memory states (a) with a corresponding electric field time profile (b).

The multiplicity of the possible remnant strains likely relates to the multiplicity of domain wall pinning in the FE domains structure,[23] and the question of scaling down the size of the device cell remains open. Scanning probe microscopy imaging of ferroelectric (also piezoelectric) substrates however provides an estimate, indicating that sub-100 nm domains structures can be stabilized and controlled in micron-sized cells.[24] Considering the difficulties in stabilizing the polarization states of FE cells when miniaturizing their sizes,[25] an approach using an analog or multi-state storage scheme can possibly remain a contender for applications with cell sizes down to a few 100s of nm.[26,27] We hope that our findings of sub-coercive realization of such states can trigger interest for applications.

As demonstrated, non-volatile on-demand switching of built-in multiple permanent states with deterministic control can be reproducibly obtained, with no experimental indications of limitations in the number of states or long-term stability of the system. This also provides a "room inside"[28,29] for ferroelectric-based memory devices operating at small electric fields with lower power consumption. The fact that internal electric field can also be changed by light in certain polar photostrictive materials: lanthanum modified ceramics PZT (Refs. 30–33) or $BiFeO_3$ (Refs. 34–37) suggests a promising optical cross-functionality at room temperature. Moreover, the possible integration of the effect reported here with spintronics should lead to an additional degree of freedom in hybrid straintronic-spintronic devices where stable changes in magnetization can be created and erased at low electric fields or light and then maintained at zero power.

Support of the Agence Nationale de la Recherche (hvSTICTSPIN ANR-13-JS04-0008-01, Labex NIE 11-LABX-0058-NIE, Investissement d'Avenir program ANR-10-IDEX-0002-02) and the technical support of the STnano cleanroom are gratefully acknowledged.

[1]E. K. H. Salje, Annu. Rev. Mater. Res. **42**, 265–283 (2012).
[2]H. Schmid, J. Phys.: Condens. Matter **20**, 434201 (2008).
[3]H. Zhang, J. Magn. Magn. Mater. **324**, 3807–3810 (2012).
[4]J. G. Wan, J.-M. Liu, G. H. Wang, and C. W. Nan, Appl. Phys. Lett. **88**, 182502 (2006).
[5]T. Zhao, S. R. Shinde, S. B. Ogale, H. Zheng, T. Venkatesan, R. Ramesh, and S. Das Sarma, Phys. Rev. Lett. **94**, 126601 (2005).
[6]I. V. Zavislyak, V. P. Sohatsky, M. A. Popov, and G. Srinivasan, Phys. Rev. B **87**, 134417 (2013).
[7]S. T. B. Goennenwein, M. Althammer C. Bihler, A. Brandlmaier, S. Geprägs, M. Opel, W. Schoch, W. Limmer, R. Gross, and M. S. Brandt, Phys. Status Solidi **2**, 96–98 (2008).
[8]C. Bihler, M. Althammer, A. Brandlmaier, S. Geprägs, M. Weiler, M. Opel, W. Schoch, W. Limmer, R. Gross, M. S. Brandt et al., Phys. Rev. B **78**, 045203 (2008).
[9]A. W. Rushforth, E. De Ranieri, J. Zemen, J. Wunderlich, K. W. Edmonds, C. S. King, E. Ahmad, R. P. Campion, C. T. Foxon, B. L. Gallagher, S. T. B. Goennenwein et al., Phys. Rev. B **78**, 085314 (2008).
[10]A. Brandlmaier, S. Geprägs, M. Weiler, A. Boger, M. Opel, H. Huebl, C. Bihler, M. S. Brandt, B. Botters, D. Grundler et al., Phys. Rev. B **77**, 104445 (2008).
[11]E. De Ranieri, A. W. Rushforth, K. Výborný, U. Rana, E. Ahmad, R. P. Campion, C. T. Foxon, B. L. Gallagher, A. C. Irvine, J. Wunderlich et al., New J. Phys. **10**, 065003 (2008).
[12]T.-K., Chung, G. P. Carman, and K. P. Mohanchandra, Appl. Phys. Lett. **92**, 112509 (2008).
[13]N. Lei, T. Devolder, G. Agnus, P. Aubert, L. Daniel, J.-V. Kim, W. Zhao, T. Trypiniotis, R. P. Cowburn, C. Chappert et al., Nature Commun. **4**, 1378 (2013).
[14]K. Roy, S. Bandyopadhyay, and J. Atulasimha, Phys. Rev. B **83**, 224412 (2011).
[15]K. Roy, S. Bandyopadhyay, and J. Atulasimha, Appl. Phys. Lett. **99**, 063108 (2011).
[16]P. Zubko, G. Catalan, and A. K. Tagantsev, Annu. Rev. Mater. Res. **43**, 387–421 (2013).
[17]H. Lu, C.-W. Bark, D. Esque de los Ojos, J. Alcala, C. B. Eom, G. Catalan, and A. Gruverman, Science **336**, 59–61 (2012).
[18]See http://www.piceramic.com/piezo_materials_1.php for commercially available (from PI) PZT ceramics were used as a substrate reference PIC 255 (PQYY + 0165) CEP2127.
[19]B. Kundys, Yu. Bukhantsev, S. Vasiliev, D. Kundys, M. Berkowski, and V. P. Dyakonov, Rev. Sci. Instrum. **75**, 2192 (2004).
[20]E. Cross, Nature **432**, 24 (2004).
[21]D. Damjanovic, *Science of Hysteresis* (Elsevier, 2006), Vol. III, pp. 337–465.
[22]E. du Trémolet de Lacheisserie, *Magnetostriction: Theory and Application of Magnetoelasticity* (CRC Press, BocaRaton, FL, 1993), p. 327.
[23]D. V. Taylor, D. Damjanovic, and N. Setter, Ferroelectrics **224**, 299 (1999).
[24]P. Paruch, T. Tybell, and J.-M. Triscone, Appl. Phys. Lett. **79**, 530 (2001).
[25]N. Balke, S. Choudhury, S. Jesse, M. Huijben, Y. H. Chu, A. P. Baddorf, L. Q. Chen, R. Ramesh, and S. V. Kalinin, Nat. Nanotechnol. **4**, 868 (2009).
[26]H. Yamada, V. Garcia, S. Fusil, S. Boyn, M. Marinova, A. Gloter, S. Xavier, J. Grollier, E. Jacquet, C. Carrétéro, C. Deranlot, M. Bibes, and A. Barthélémy, ACS Nano **7**, 5385 (2013).
[27]S. Boyn, S. Girod, V. Garcia, S. Fusil, S. Xavier, C. Deranlot, H. Yamada, C. Carrétéro, E. Jacquet, M. Bibes et al., Appl. Phys. Lett. **104**, 052909 (2014).
[28]V. Garcia and M. Bibes, Nature **483**, 279 (2012).
[29]D. Lee, S. M. Yang, T. H. Kim, B. C. Jeon, Y. S. Kim, J.-G. Yoon, H. N. Lee, S. H. Baek, C. Beom Eom, and T. W. Noh, Adv. Mater. **24**, 402–406 (2012).
[30]K. Takagi, S. Kikuchi, J.-F. Li, H. Okamura, R. Watanabe, and A. Kawasaki, J. Am. Ceram. Soc. **87**(8), 1477 (2004).
[31]K. Uchino, M. Aizawa, and S. Nomura, Ferroelectrics **64**, 199 (1985).
[32]T. Sada, M. Inoue, and K. Uchino, J. Ceram. Soc. Jpn. **95**, 499 (1987).
[33]S.-Y. Chu and K. Uchino, Ferroelectrics **174**, 185 (1995).
[34]B. Kundys, M. Viret, D. Colson, and D. O. Kundys, Nature Mater. **9**, 803 (2010).
[35]B. Kundys, C. Meny, M. R. J. Gibbs, V. Da Costa, M. Viret, M. Acosta, D. Colson, and B. Doudin, Appl. Phys. Lett. **100**, 262411 (2012).
[36]D. Schick, M. Herzog, H. Wen, P. Chen, C. Adamo, P. Gaal, D. G. Schlom, P. G. Evans, Y. Li, M. Bargheer et al., Phys. Rev. Lett. **112**, 097602 (2014).
[37]H. Wen, Phys. Rev. Lett. **110**, 037601 (2013).